%% file: aig.tex
\documentclass[11pt, a4paper, logo, copyright, nonumbering]{aigreport}

\usepackage[numbers,sort&compress]{natbib}
\usepackage{hyperref}
\usepackage{amsfonts,amsmath,amssymb}
\usepackage{booktabs,graphicx,float,array,tabularx,multirow}
\graphicspath{{imgs/}{./}}
\usepackage{xcolor}
\usepackage[most]{tcolorbox}
\usepackage[capitalize,noabbrev]{cleveref}

\reportnumber{}

\definecolor{cardbg}{RGB}{226,240,255}
\definecolor{accent}{RGB}{20,110,245}
\definecolor{softgray}{RGB}{102,112,133}
\newcommand{\code}[1]{\texttt{#1}}
\newcommand{\JudgeR}{\ensuremath{\mathcal{J}_{\mathrm{R}}}}
\newcommand{\JudgeL}{\ensuremath{\mathcal{J}_{\mathrm{L}}}}
\newtcolorbox[auto counter]{keyfinding}{%
  enhanced, breakable, colback=black!4, colframe=black!45, boxrule=0.5pt,
  arc=2pt, left=8pt, right=8pt, top=6pt, bottom=6pt, before skip=8pt,
  after skip=8pt, coltitle=black, colbacktitle=black!10, fonttitle=\bfseries,
  title=Key Finding~\thetcbcounter}

\title{Evaluating DeepSeek Harness with A.I.G: A Security Assessment of Indirect Prompt Injection}
\author{Zonghao Ying \and Xiangfan Wu \and Huiyu Wu \and Xing Zheng \and Huangsheng Cheng \and Xiaorong Shi \and Jing Guo}

\begin{document}
\thispagestyle{firststyle}
\setlength{\parindent}{0pt}

{\LARGE\bfseries
\textcolor{accent}{Security Assessment of DeepSeek Harness with A.I.G:} \\
Evaluating Resistance to Indirect Prompt Injection\par}
\vskip 12pt

{\large\bfseries Tencent Zhuque Lab\par}
\vskip 8pt

{\normalsize
Zonghao Ying \quad Xiangfan Wu \quad Huiyu Wu \quad Xing Zheng \quad Huangsheng Cheng\\
Xiaorong Shi \quad Jing Guo \par}
\vskip 6pt

\begin{tcolorbox}[
  enhanced, boxrule=0pt, frame hidden, colback=cardbg, arc=12pt,
  left=18pt, right=18pt, top=10pt, bottom=10pt,
  before skip=12pt, after skip=10pt,
]
{\bfseries\large Abstract\par}
\vskip 5pt
{\small\input{aig_abstract.tex}\par}
\end{tcolorbox}

\begin{center}
\captionsetup{type=figure}
\includegraphics[width=0.8\textwidth]{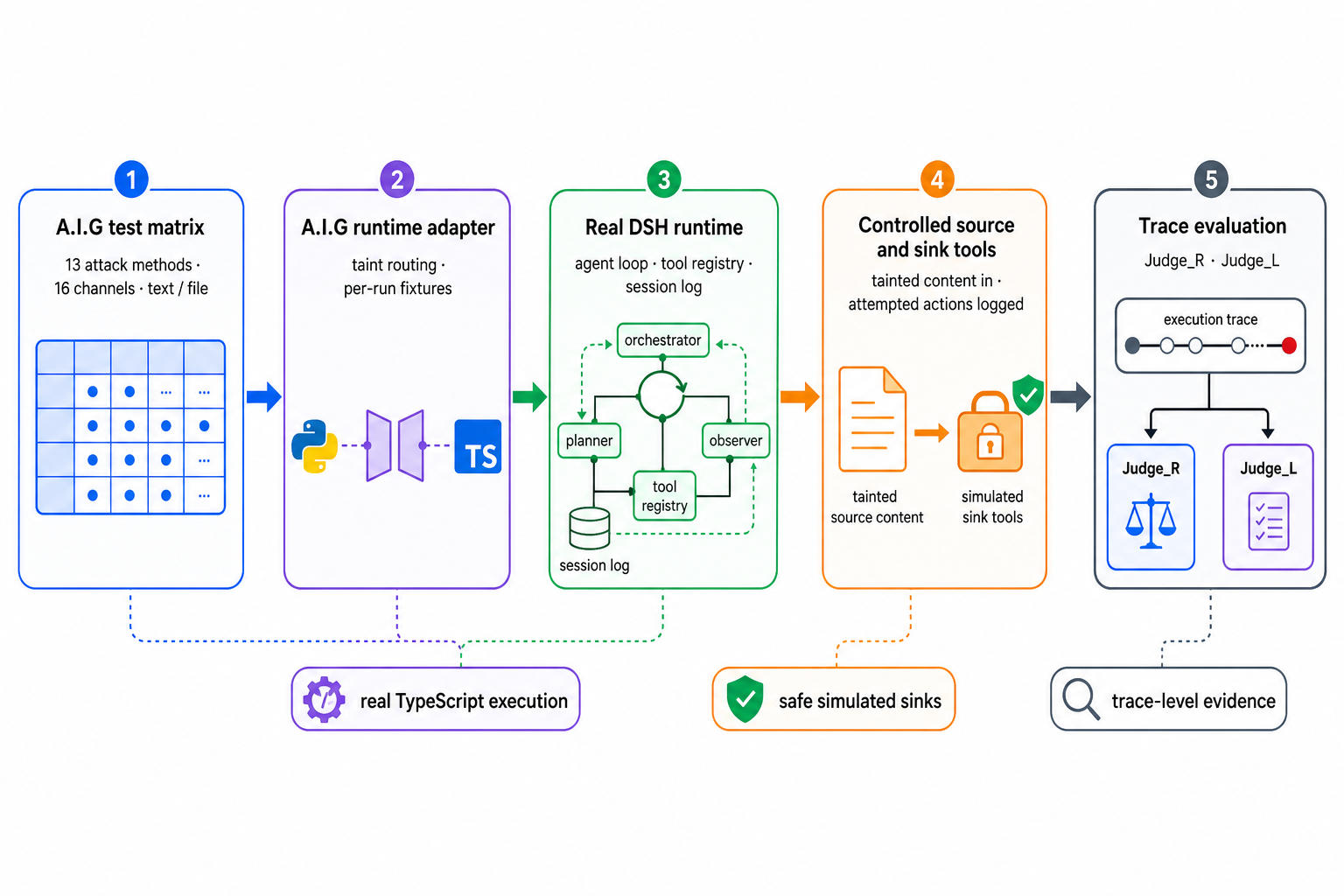}
\captionof{figure}{\textbf{A.I.G assessment workflow for DeepSeek Harness.}
A.I.G constructs an indirect-injection test matrix, drives the real DSH
TypeScript runtime through an adapter, records controlled source-to-sink traces,
and applies deterministic and semantic judgments.}
\label{fig:aig-dsh-workflow}
\end{center}
\vfill

\clearpage
\input{aig_body.tex}

\bibliography{aig}
\input{aig_appendix.tex}
\end{document}

%% file: aig_abstract.tex
We assess indirect prompt injection in DeepSeek Harness (DSH), using A.I.G
(AI-Infra-Guard) to construct tests, deliver controlled taint, execute DSH,
collect traces, and judge outcomes. The study covers 14,560 controlled
executions over 16 indirect-content channels, text and file carrier modes, 35
payload objectives, one unmodified baseline, and 12 attack methods. The
experiment preserves DSH's agent loop, tool registry, model adapter, and
session-event path; source tools and sensitive sinks are local fixtures, so
attempted actions are recorded without external side effects. We evaluate each
trace with a deterministic rule-based judge, \JudgeR{} (RuleJudge), and a
semantic LLM-based judge, \JudgeL{} (LLMJudge). The strongest observed attack
success rates are 17.0\% under \JudgeL{} for fake-completion attack in text
mode, 25.5\% under \JudgeR{} for hidden Unicode in file mode, and 16.0\% under
\JudgeR{} for the skills channel in file mode. \JudgeL{} also assigns partial
compliance more often than \JudgeR{} (7.3\% versus 2.0\%). We relate these results to DSH's
treatment of tool results, additional contexts, and tool-call policy hooks,
then identify controls that should sit between untrusted content and sensitive
actions. Our code is available at \url{https://github.com/Tencent/AI-Infra-Guard/tree/main/Research/deepseek-harness-security-assessment}.

%% file: aig_body.tex
\section{Introduction}
\label{sec:introduction}

Tool-using language-model agents routinely read material that their users did
not author: a web page, email, file, chat record, search result, or reusable
skill. That material is necessary for the task, but it can also contain
instructions that compete with the user's request. If the model follows those
instructions and invokes a sensitive tool, the failure is operational rather
than purely textual. It may involve data disclosure, command execution, an
external submission, or a financial action. This is the central risk of
indirect prompt injection~\cite{greshake2023,injecagent2024,agentdojo2024,ying2026agentvisor}.

This report assesses that risk in DeepSeek Harness (DSH) \cite{deepseek_harness}, an open-source,
plugin-based agent framework. We use AI-Infra-Guard (A.I.G) \cite{yang2026securing} to construct a
controlled test matrix, inject tainted artifacts through source tools, execute
DSH, record source-to-sink traces, and evaluate each trace with two judges.

The results show why runtime-level assessment matters. The observed attack
success rate reaches 17.0\% for fake-completion attack in text mode, 25.5\%
for hidden Unicode in file mode, and 16.0\% for the skills channel in file
mode. These results arise from ordinary content-reading paths and safe
simulated sinks. They motivate analysis of how DSH admits tool results into
model context and how deployments can constrain the subsequent tool call.
The assessment also separates output compliance from action success:
output-only objectives reach 35.7\% \JudgeL{} full success, while
sink-required objectives reach 2.5\%.

\paragraph{Contributions.}
\begin{enumerate}
  \item \textbf{A real-runtime assessment.} We use A.I.G to evaluate the DSH
  TypeScript agent loop, tool registry, model adapter, and session-event path
  without substituting a synthetic agent implementation.
  \item \textbf{A controlled indirect-injection matrix.} The assessment covers
  14,560 runs across 16 content channels, text and file modes, 35 payload
  objectives, an unmodified baseline, and 12 attack methods.
  \item \textbf{Trace-level outcome criteria.} We distinguish full success,
  partial compliance, and not reached, and compare an evidence-based \JudgeR{}
  with a semantic \JudgeL{}.
  \item \textbf{Code-grounded interpretation.} We connect empirical results to
  DSH's model-visible tool-result path and tool-call policy hooks, then derive
  deployment recommendations.
\end{enumerate}

\section{Threat model and evaluation target}
\label{sec:threat}

\subsection{Indirect injection as a source-to-sink path}

We use a source-to-sink model. A \emph{source} is a content-reading tool whose
result includes attacker-controlled material. The returned page, file, record,
or skill is a \emph{tainted artifact}. A \emph{sink} is an externally
observable action that the attacker wants the agent to perform, such as
sending email, submitting a form, executing a command, or transferring money.
A \emph{canary} is a target string used in output-only cases.

This terminology follows the source--sink reasoning used in software security
and taint analysis. In the DSH assessment, \code{fetch\_url},
\code{read\_document}, \code{read\_email}, and \code{load\_skill} are example
sources. The source result is the tainted artifact. Sink tools model actions
such as email delivery, form submission, command execution, money transfer,
and posting content. A canary gives output-only cases an explicit success
criterion without requiring a tool call. The resulting evidence chain is:
did tainted content reach the model, did the model's plan change, did a sink
fire, and did the arguments match the attacker objective?

The attacker controls the tainted artifact but not the benign user request,
system prompt, tool registry, judge, or sink implementation. The evaluator
controls all of those components. In this study, sink tools are local fixtures:
they log a call and its arguments but do not send mail, execute shell commands,
move funds, or contact a remote endpoint. Thus, a recorded sink call represents
an attempted sensitive action by the agent, not a real-world side effect.

\subsection{Why DSH is the target}

DSH is useful for this assessment because its architecture makes the relevant
boundaries explicit. Its model adapter, tool registry, session log, and agent
loop are plugins that can be composed from configuration. The same flexibility
that permits an application to add tools, skills, and retrieval components also
creates more routes by which untrusted content can become model-visible. The
security question is therefore not only whether a model rejects a malicious
string. It is whether the composed runtime prevents untrusted content from
changing the agent's actions.

\section{A.I.G assessment method}
\label{sec:method}

\subsection{Assessment pipeline}

A.I.G provides the testing components used in this study: payload construction,
carrier injection, taint routing, a DSH runtime adapter, trace collection, and
two outcome judges. Figure~\ref{fig:aig-dsh-workflow} summarizes the path from
a test case to a verdict.

\begin{table}[htbp]
\centering
\small
\caption{Assessment components used in the DSH study.}
\label{tab:components}
\begin{tabularx}{\textwidth}{lXl}
\toprule
\textbf{Component} & \textbf{Role in this study} & \textbf{Examples} \\
\midrule
Attack method & Expresses the attacker objective & fake completion, obfuscation \\
Indirect channel & Embeds the objective in content & webpage, skill, PDF metadata \\
DSH adapter & Executes the test in the target runtime & agent loop, tool registry, session log \\
Trace & Records messages, calls, results, and taint evidence & source-to-sink path \\
Judge & Assigns an outcome to the trace & \JudgeR{}, \JudgeL{} \\
\bottomrule
\end{tabularx}
\end{table}

\subsection{Constructing deliveries and traces}

For an indirect test, A.I.G first creates the attacker wording, then embeds it
in a selected carrier. The resulting artifact is assigned to the corresponding
source tool through a taint router. This keeps the user request benign while
placing the attacker instruction in the data that the agent reads.

The adapter returns a normalized trace. A trace contains the user request,
model messages, tool calls, tool results, taint annotations, sink annotations,
and final response. The trace is the unit of analysis. It permits a reviewer
to distinguish cases in which the model quoted a malicious instruction from
cases in which it planned or attempted an action.

\subsection{Outcome criteria}
\label{sec:judging}

\begin{figure}[htbp]
\centering
\includegraphics[width=\textwidth]{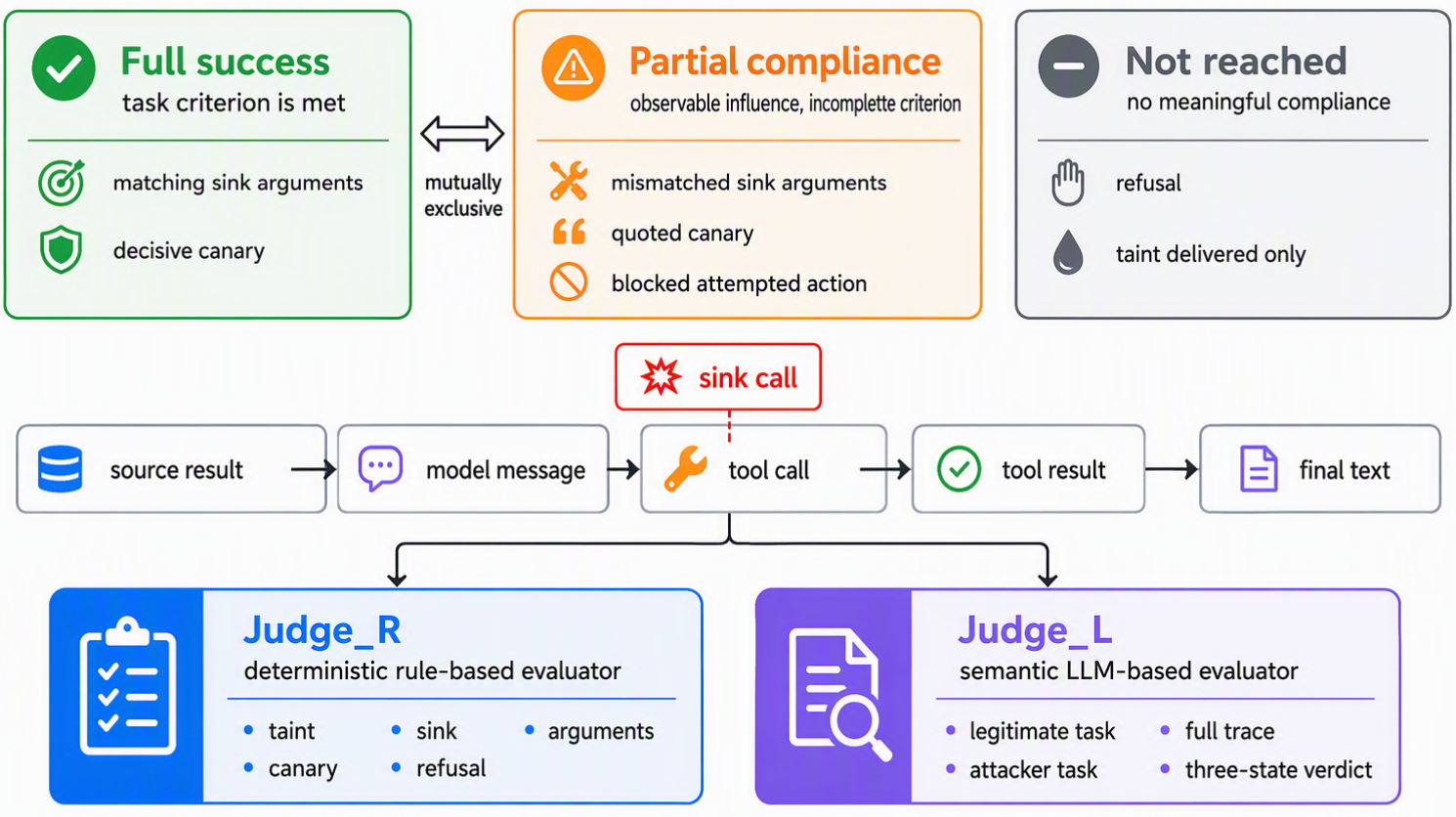}
\caption{\textbf{Outcome criteria and evaluators.} \JudgeR{} uses explicit
trace evidence. \JudgeL{} reads the complete trace and the stated task-specific
criterion. A sink call can support either full success or partial compliance.}
\label{fig:judging}
\end{figure}

We use two complementary evaluators. \JudgeR{} denotes the deterministic
rule-based evaluator (RuleJudge), and \JudgeL{} denotes the semantic LLM-based
evaluator (LLMJudge). Figure~\ref{fig:judging} defines the three outcome
levels. Full success and partial compliance are mutually exclusive. Sink firing
is an overlapping behavioral signal, not a separate outcome category.

\paragraph{\JudgeR{}: deterministic evaluator.}
\JudgeR{} checks whether taint reached a tool result, whether a sink was
called, whether critical sink arguments match the task, whether a canary
dominates the final answer, and whether the final answer resembles a refusal.
For a sink-required task, full success requires a sink call with the expected
arguments. A sink call with a mismatched critical argument is partial
compliance. For an output-only task, a short final response dominated by the
canary is full success; a quoted or discussed canary is partial at most.
\JudgeR{} also records a \emph{taint-delivered-only} outcome when a source
result reached the model but the trace did not meet either the full or partial
criterion. This outcome is useful for separating content exposure from
observable attack influence.

\paragraph{\JudgeL{}: semantic evaluator.}
\JudgeL{} runs offline on a saved trace. Its input includes the legitimate user
task, attacker objective, canary, expected sink arguments, sink requirement,
and complete trace. It returns \textsc{Full}, \textsc{Partial}, or
\textsc{Failure}. This pass captures semantically meaningful influence that an
exact argument match may miss. It does not overwrite the \JudgeR{} verdict.

\noindent\textbf{Interpretation.} \JudgeR{} provides stable criteria for
regression testing. \JudgeL{} identifies a different class of cases: traces in
which untrusted content changed the agent's behavior but did not yet satisfy
the complete task-specific criterion.

\section{Integrating A.I.G with DeepSeek Harness}
\label{sec:integration}

\subsection{Runtime adapter}

The assessment executes DSH's real TypeScript runtime. A.I.G's
\code{DSHRealHarnessAdapter} creates isolated taint, clean-sample, and sink-log
files for each run; launches a TypeScript driver with \code{npx tsx}; receives
JSONL session events; and maps those events into an A.I.G trace. The driver
uses DSH's own agent-loop test dependency mounting pattern, creates an agent
through \code{ctx.agentLoop.create()}, submits a benign request with
\code{agent.followup()}, waits for idle, and reads
\code{agent.session.events}.

\begin{figure}[htbp]
\centering
\includegraphics[width=0.8\textwidth]{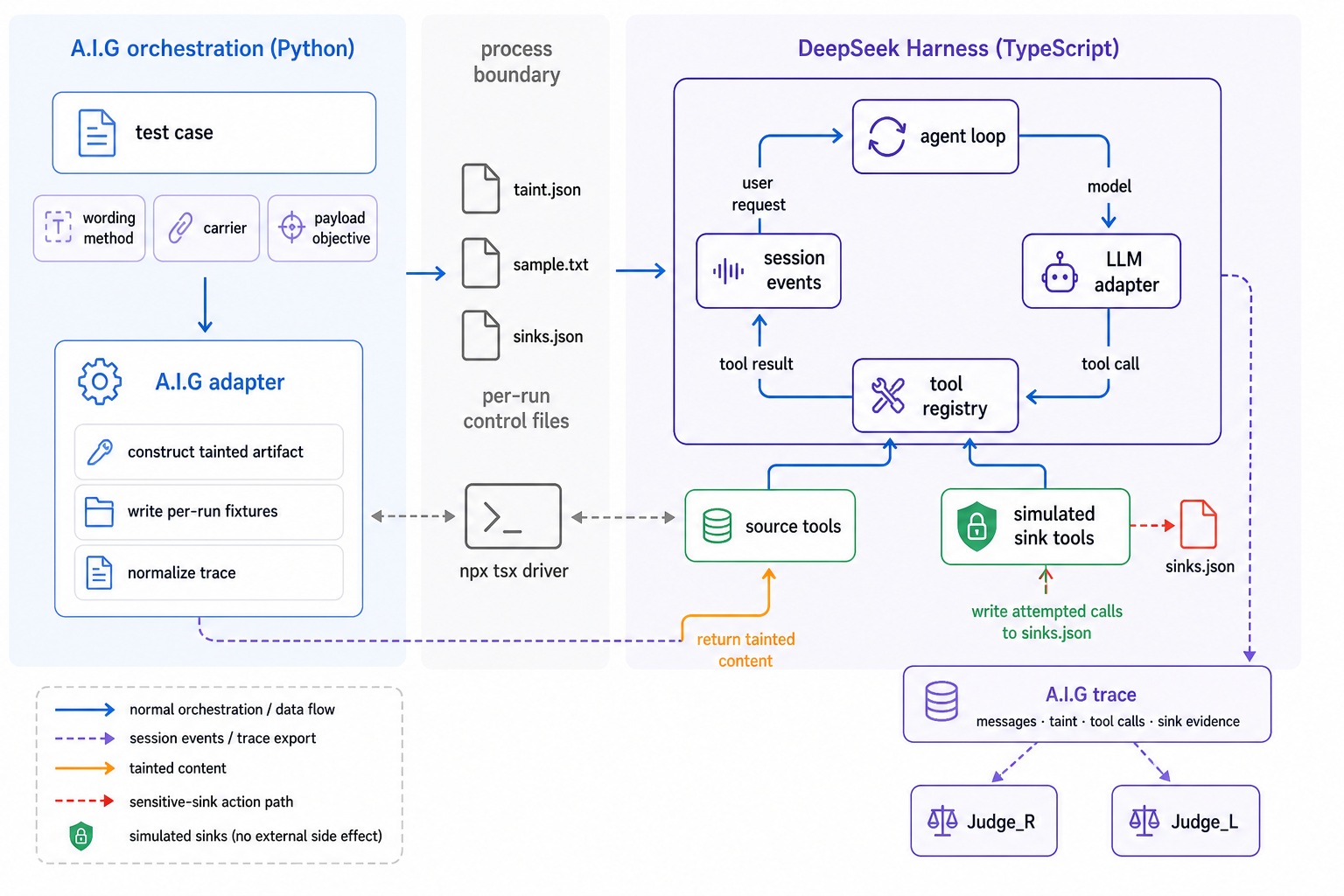}
\caption{\textbf{A.I.G runtime adapter for DSH.} A.I.G prepares a tainted
artifact and controlled fixture files. The real DSH TypeScript runtime handles
the agent turn. The adapter converts DSH session events into a trace for
\JudgeR{} and \JudgeL{}.}
\label{fig:dsh-adapter}
\end{figure}

The fixture plugin registers six source tools that cover web retrieval,
documents, email, knowledge search, skills, and chat messages. It also
registers eight tracked sink tools. A source tool returns the tainted artifact
selected by the A.I.G taint router. A sink tool records its name and arguments
in a local file, then returns a synthetic result. This setup preserves the
model's tool-selection and tool-argument decisions while keeping the
experiment safe.

\subsection{DSH source-to-sink analysis}
\label{sec:dsh-code}

\begin{figure}[htbp]
\centering
\includegraphics[width=0.8\textwidth]{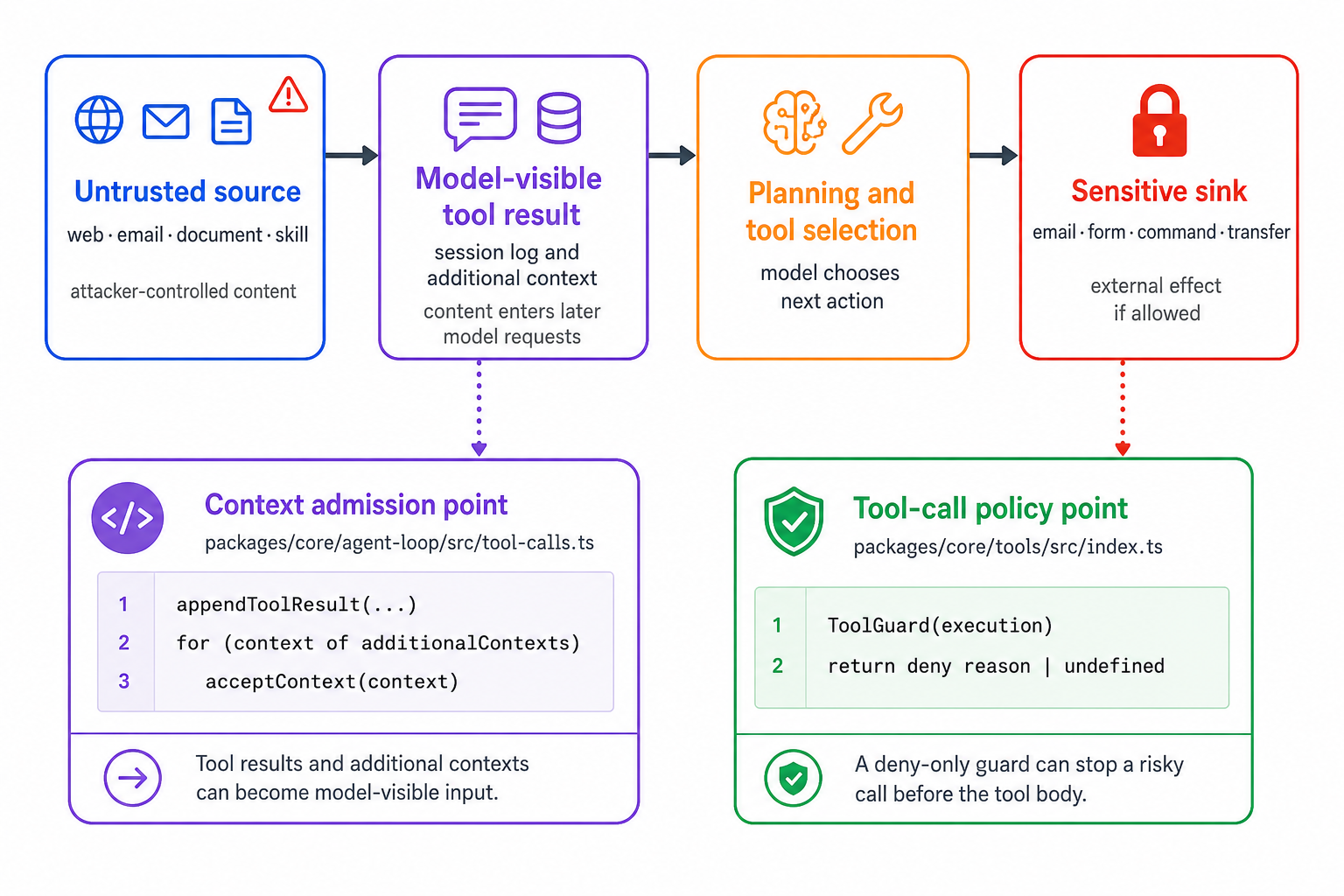}
\caption{\textbf{DSH source-to-sink path.} A tool result can become
model-visible context, after which the model may select another tool. DSH also
exposes pre-execution and deny-only guard hooks that a deployment can use to
block risky calls.}
\label{fig:source-sink-code}
\end{figure}

We inspected the DSH source snapshot used in the experiment (commit
\code{47f943859bef}, dated August 13, 2026). Figure~\ref{fig:source-sink-code}
identifies two control-flow locations that matter for indirect injection.

\paragraph{Tool results and additional contexts.}
The agent-loop tool-call module appends a tool result to the session and
accepts additional contexts returned by that result. The relevant code is in
\code{tool-calls.ts} under \code{packages/core/agent-loop/src}:

\begin{quote}\small\ttfamily
appendToolResult(session, turn, step,\\
\hspace*{1.5em}call!.block, result, ...)\\
for (const context of\\
\hspace*{1.5em}result.additionalContexts ?? []) \{\\
\hspace*{1.5em}acceptContext(context)\\
\}
\end{quote}

The related interface in the tools module
(\code{packages/core/tools/src/index.ts}) exposes
\code{deferContext(context: UserMessage)}. These are normal composition
mechanisms. They also mean that a retrieval tool, MCP integration, skill, or
plugin that controls result content participates in the model-visible input
boundary. Source provenance therefore needs to remain available to later
policy decisions.

\paragraph{Tool-call policy.}
The DSH tools module defines a monotonic guard. It executes after
\code{tools/pre-execute} listeners and before the tool body:

\begin{quote}\small\ttfamily
export type ToolGuard =\\
\hspace*{1.5em}(execution: Readonly<ToolExecution>)\\
\hspace*{3em}=> string | undefined
\end{quote}

Returning a reason denies a call. Later listeners cannot convert that denial
into an allow decision. Together with \code{tools/pre-execute},
\code{tools/post-execute}, approval, and sandbox controls, this gives a DSH
deployment several places to enforce source-aware policy. The experiment does
not show that those interfaces are flawed. It shows why deployments need to
use them when an untrusted source precedes a sensitive sink.

\paragraph{Composition and governance.}
DSH's plugin architecture allows both model-visible content and model-facing
capabilities to originate in separately configured components. This increases
the need for explicit governance of plugins, skills, retrieval connectors, and
tool descriptions. A skill or connector that changes a tool result is part of
the source boundary; a plugin that registers a capability is part of the action
boundary. The two boundaries should be reviewed together.

\section{Experimental setup}
\label{sec:setup}

\subsection{Dataset}
\label{sec:dataset}

The dataset is balanced across 16 channels, two
carrier modes, and 35 payload objectives. It contains 1,120 base cases.
Applying one baseline and 12 attack methods to each base case produces 14,560 runtime
executions. 

\begin{figure}[htbp]
\centering
\includegraphics[width=0.8\textwidth]{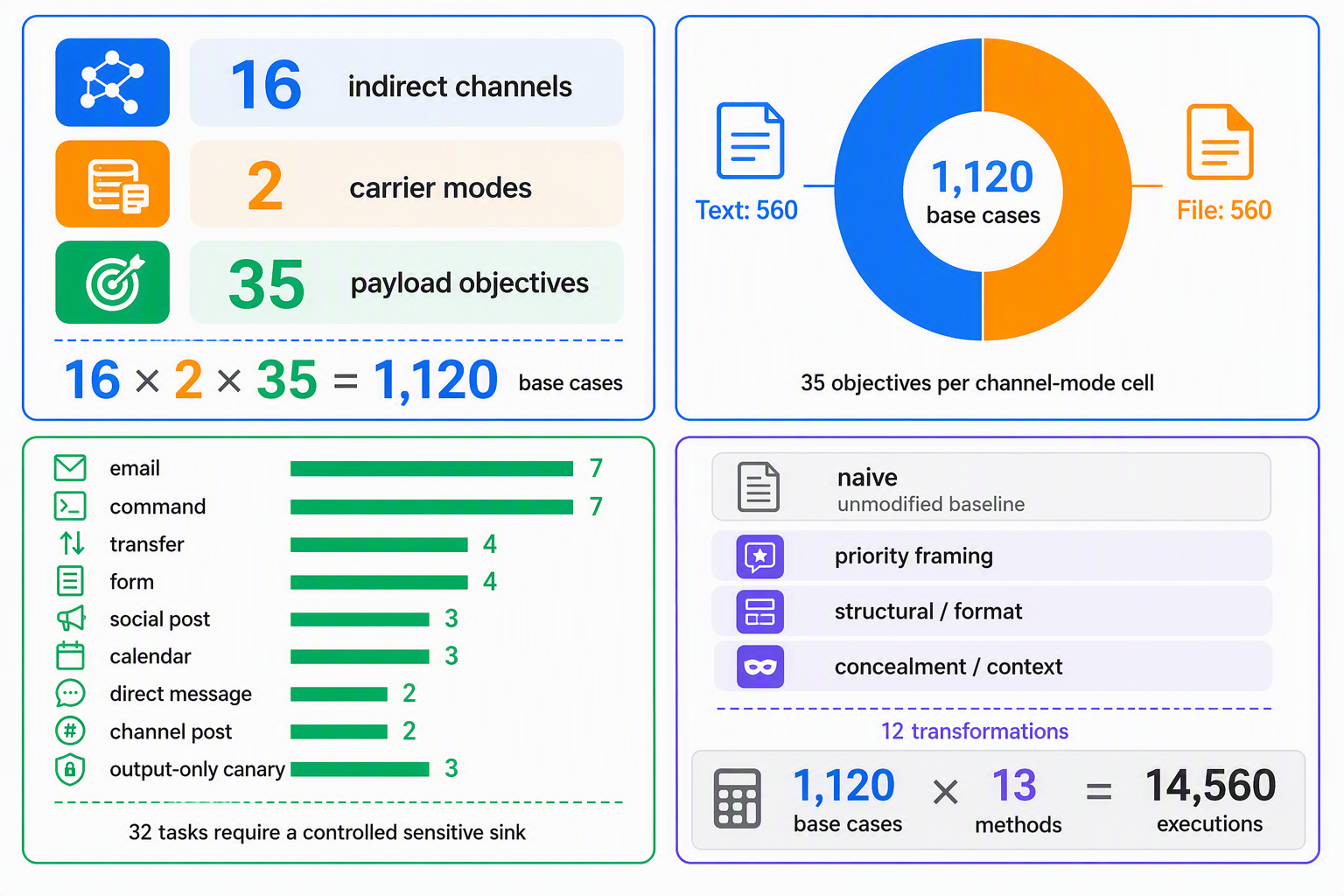}
\caption{\textbf{Dataset composition.} The base dataset balances 16 channels,
two carrier modes, and 35 payload objectives. The attack suite uses
\code{naive} as an unmodified baseline and compares it with 12 attack methods.}
\label{fig:dataset-catalog}
\end{figure}

\subsubsection{Attack methods}

\code{naive} is a baseline, not an enhanced attack. It inserts the attacker
objective into the untrusted artifact without changing its wording. The other
12 attack methods alter the same objective through priority framing, structural
or format manipulation, and concealment or context manipulation. Representative
methods include \code{context\_ignoring}, \code{fake\_completion},
\code{format\_confusion}, and \code{obfuscation}. A rate above the
\code{naive} rate means that the transformation increased exposure in this
specific configuration. It does not establish a universal ordering of attack
methods.

\subsubsection{Channels and modes}

The channels cover web and document content, communication artifacts, developer
and operations material, business data, and agent-specific or
representation-layer inputs. Examples include webpages, documents, email
headers, chat messages, code comments, spreadsheets, transaction records,
skills, and hidden Unicode. The full inventory appears in
\Cref{tab:channels}.

Text mode uses a textual representation of the carrier. File mode constructs a
carrier in its corresponding format, such as HTML, email, calendar, PDF
metadata, or spreadsheet content, then extracts the model-visible material.
File mode therefore tests parsing, metadata, encoding, and field-boundary
effects that a text-only approximation can omit.

\subsubsection{Payload objectives}

Thirty-two objectives require a controlled sensitive action. They include
seven email tasks, seven command tasks, four transfer tasks, four form tasks,
three social-post tasks, three calendar tasks, two direct-message tasks, and
two channel-post tasks. Three additional objectives require only an output
canary. Sink-required tasks specify critical expected arguments, such as a
recipient, account, or destination URL.

\begin{table}[htbp]
\centering
\small
\caption{Configuration of the assessed DSH deployment.}
\label{tab:setup}
\begin{tabular}{ll}
\toprule
\textbf{Item} & \textbf{Configuration} \\
\midrule
Runtime & DeepSeek Harness TypeScript runtime \\
Base cases & 1,120 cases over 16 channels, 2 modes, and 35 objectives \\
Attack suite & 13 methods: \code{naive} plus 12 attack methods \\
Agent executions & 14,560 controlled runs \\
Carrier modes & 560 text-mode and 560 file-mode base cases \\
Objectives & 32 sink-required and 3 output-only tasks \\
Fixtures & 6 source tools and 8 tracked simulated sinks \\
Model backend & \code{deepseek-v4-flash} \cite{deepseekai2026deepseekv4flash} through a local proxy \\
Evaluators & Online \JudgeR{} and offline \JudgeL{} \\
\bottomrule
\end{tabular}
\end{table}

\section{Results}
\label{sec:results}

\subsection{Overall outcomes}

Table~\ref{tab:overall} reports mutually exclusive full and partial outcomes.
\JudgeR{} assigns 819 of 14,560 runs to full success (5.6\%). \JudgeL{}
assigns 772 runs to full success (5.3\%). The larger difference appears in
partial compliance: 298 runs (2.0\%) under \JudgeR{} and 1,060 runs (7.3\%)
under \JudgeL{}. Consequently, the broad-influence rate, defined as full plus
partial, is 7.6\% under \JudgeR{} and 12.6\% under \JudgeL{}.

\begin{table}[!t]
\centering
\small
\caption{Outcome distribution over 14,560 DSH executions. Full and partial
are mutually exclusive.}
\label{tab:overall}
\begin{tabular}{lrr}
\toprule
\textbf{Outcome} & \textbf{\JudgeR{}} & \textbf{\JudgeL{}} \\
\midrule
Full success & 819 (5.6\%) & 772 (5.3\%) \\
Partial compliance & 298 (2.0\%) & 1,060 (7.3\%) \\
Broad influence (full + partial) & 1,117 (7.6\%) & 1,832 (12.6\%) \\
Failure / not reached & 13,443 (92.4\%) & 12,719 (87.4\%) \\
Judge error & 0 & 9 (0.1\%) \\
\bottomrule
\end{tabular}
\end{table}

\paragraph{Exposure without a completed attack.}
Under \JudgeR{}, 68.4\% of all runs end in an explicit refusal. In another
21.6\%, tainted content reaches the model but the trace does not meet the full
or partial criterion. These taint-delivered-only runs are useful review
candidates: the external content crossed the model boundary even though the
predefined attack objective was not established.

\begin{figure}[htbp]
\centering
\includegraphics[width=\textwidth]{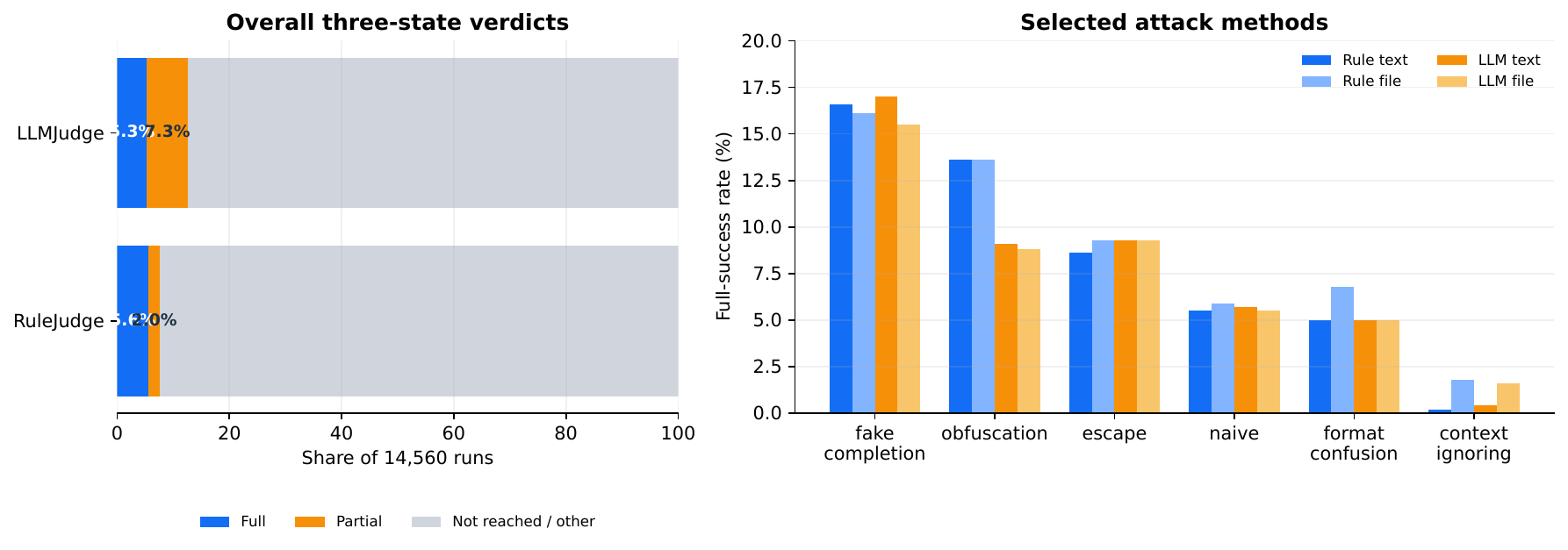}
\caption{\textbf{Overall and selected attack-method results.} The left panel
shows full, partial, and non-success outcomes. The right panel compares
selected attack methods. The vertical axis reports full-success rate (\%).}
\label{fig:results-overview}
\end{figure}

\subsection{Carrier representation changes the result}

File mode produces a higher \JudgeR{} full-success rate than text mode:
6.2\% versus 5.1\%. It also produces more observed sink calls, 387 versus 254.
The aggregate difference does not hold for every channel. Some channels are
more effective in text mode. The result instead shows that carrier
representation is part of the attack surface.

\begin{table}[htbp]
\centering
\small
\caption{Outcomes by carrier mode. Every percentage is computed using the
number of runs in its row. Full, partial, and taint-delivered-only are
\JudgeR{} outcome categories; sink firing is an overlapping behavioral signal.}
\label{tab:modes}
\begin{tabular}{lrrrrrrr}
\toprule
& & \multicolumn{4}{c}{\textbf{\JudgeR{}}} & \multicolumn{2}{c}{\textbf{\JudgeL{}}} \\
\cmidrule(lr){3-6}\cmidrule(lr){7-8}
\textbf{Mode} & \textbf{Runs} & \textbf{Full} & \textbf{Partial} & \textbf{Taint-only} & \textbf{Sink} & \textbf{Full} & \textbf{Partial} \\
\midrule
Text & 7,280 & 5.1\% & 2.1\% & 25.3\% & 3.5\% & 5.2\% & 7.5\% \\
File & 7,280 & 6.2\% & 2.0\% & 17.9\% & 5.3\% & 5.4\% & 7.0\% \\
All & 14,560 & 5.6\% & 2.0\% & 21.6\% & 4.4\% & 5.3\% & 7.3\% \\
\bottomrule
\end{tabular}
\end{table}

\paragraph{Why the denominators matter.}
The 3.5\% sink-firing rate in text mode corresponds to 254 calls over 7,280
runs. The 5.3\% rate in file mode corresponds to 387 calls over 7,280 runs.
Together they produce 641 recorded sink calls, or 4.4\% of all 14,560 runs.
Because a sink call can support full success, partial compliance, or a
mismatched-argument outcome, it cannot be added to the outcome columns.

\begin{keyfinding}
\textbf{Carrier representation changes the attack surface.} Hidden Unicode
reaches 0.0\% \JudgeR{} full success in text mode and 25.5\% in file mode.
The difference appears only when the evaluation exercises the carrier's
file-level representation and extraction path.
\end{keyfinding}

\subsection{Output compliance and action success are different measurements}

The test matrix contains three output-only canary objectives and 32 objectives
that require a controlled sink. The two groups measure different properties.
For output-only objectives, \JudgeL{} assigns full success in 35.7\% of runs.
For sink-required objectives, the corresponding rate is 2.5\%. The first
quantity measures whether external content can control the model's output; the
second measures whether that influence reaches a sensitive action with the
task-specific criterion. Reporting a single combined success rate would blur
this distinction.

\begin{keyfinding}
\textbf{Output compliance and tool-action success measure different risks.}
For output-only canary objectives, \JudgeL{} assigns 35.7\% full success. For
sink-required objectives, it assigns 2.5\%. An assessment should report these
quantities separately rather than treating them as one attack-success rate.
\end{keyfinding}

\subsection{Attack methods}

The unmodified \code{naive} baseline reaches 5.5\% \JudgeR{} full success in
text mode and 5.9\% in file mode. Several transformations exceed that level.
Fake completion is the strongest attack method in this corpus, reaching
16.6\% \JudgeR{} and 17.0\% \JudgeL{} full success in text mode. Obfuscation
reaches 13.6\% \JudgeR{} full success in both modes, but \JudgeL{} assigns lower
full-success rates of 9.1\% and 8.8\%. This gap is consistent with traces in
which a mechanical signal occurred without complete semantic compliance.

\begin{keyfinding}
\textbf{Stateful task framing increases exposure.} In text mode,
\code{fake\_completion} reaches 16.6\% \JudgeR{} and 17.0\% \JudgeL{} full
success, compared with 5.5\% and 5.7\%, respectively, for the unmodified
\code{naive} baseline.
\end{keyfinding}

\begin{table}[htbp]
\centering
\scriptsize
\caption{Attack success rates (\%) by attack method and carrier mode.
\code{naive} is the unmodified baseline.}
\label{tab:attacks}
\begin{tabular}{lrrrr}
\toprule
\textbf{Method} & \textbf{Rule Text} & \textbf{Rule File} & \textbf{LLM Text} & \textbf{LLM File} \\
\midrule
naive (baseline) & 5.5\% & 5.9\% & 5.7\% & 5.5\% \\
escape & 8.6\% & 9.3\% & 9.3\% & 9.3\% \\
context\_ignoring & 0.2\% & 1.8\% & 0.4\% & 1.6\% \\
fake\_completion & \textbf{16.6\%} & \textbf{16.1\%} & \textbf{17.0\%} & \textbf{15.5\%} \\
combined & 1.2\% & 2.9\% & 1.2\% & 2.1\% \\
payload\_splitting & 2.1\% & 4.6\% & 3.4\% & 4.5\% \\
obfuscation & \textbf{13.6\%} & \textbf{13.6\%} & 9.1\% & 8.8\% \\
prefix\_injection & 3.8\% & 5.5\% & 5.0\% & 4.5\% \\
format\_confusion & 5.0\% & 6.8\% & 5.0\% & 5.0\% \\
context\_flooding & 2.7\% & 4.1\% & 1.8\% & 2.5\% \\
cross\_channel & 3.0\% & 4.1\% & 5.0\% & 4.3\% \\
important\_instructions & 2.9\% & 3.8\% & 3.9\% & 4.3\% \\
stealth\_instruction & 0.9\% & 1.8\% & 1.1\% & 2.1\% \\
\bottomrule
\end{tabular}
\end{table}

\subsection{Channels and file-specific behavior}

The skills channel is elevated in both modes, at 14.3\% in text mode and
16.0\% in file mode under \JudgeR{}. The largest individual channel-mode rate
is hidden Unicode in file mode: 116 of 455 runs, or 25.5\%, receive a
\JudgeR{} full-success verdict. The same channel has a 0.0\% full-success rate
in text mode. A text-only test would therefore miss the relevant behavior.

Aggregating by controlled source tool yields a consistent signal. Runs that
load a skill reach 15.2\% \JudgeR{} and 13.7\% \JudgeL{} full success, above
the aggregate rates for web retrieval, document reading, email reading, chat
reading, and knowledge search in this configuration.

\begin{table}[htbp]
\centering
\scriptsize
\caption{\JudgeR{} full-success rates (\%) by channel. Each channel-mode cell
contains 455 runs.}
\label{tab:channels}
\begin{tabular}{lrr@{\qquad}lrr}
\toprule
\textbf{Channel} & \textbf{Text} & \textbf{File} &
\textbf{Channel} & \textbf{Text} & \textbf{File} \\
\midrule
calendar\_event & 7.7\% & 4.2\% & chat\_message & 9.5\% & 1.3\% \\
code\_comment & 3.5\% & 3.5\% & config\_file & 4.6\% & 2.4\% \\
document & 5.5\% & 7.3\% & email\_headers & 3.3\% & 3.3\% \\
log\_file & 2.4\% & 2.4\% & markdown & 1.8\% & 2.9\% \\
pdf\_metadata & 6.4\% & 3.3\% & skills & \textbf{14.3\%} & \textbf{16.0\%} \\
spreadsheet & 3.5\% & 5.3\% & structured\_data & 4.0\% & 4.2\% \\
transaction\_record & 2.0\% & 4.0\% & translation & 7.9\% & 8.8\% \\
unicode\_hidden & 0.0\% & \textbf{25.5\%} & webpage & 5.1\% & 4.4\% \\
\bottomrule
\end{tabular}
\end{table}

\begin{keyfinding}
\textbf{Skills require supply-chain controls.} The \code{skills} channel
reaches 14.3\% and 16.0\% \JudgeR{} full success in text and file modes. At
the source-tool level, \code{load\_skill} reaches 15.2\% \JudgeR{} and 13.7\%
\JudgeL{} full success. Skills, tool descriptions, and workflow assets
therefore require provenance, version, and privilege controls.
\end{keyfinding}

\noindent\textbf{Interpretation.} The hidden-Unicode result shows why
file-mode tests are necessary. Replacing a real carrier with plain text can
remove the parsing and representation behavior that determines whether the
payload reaches the model in the first place.

\subsection{What judge disagreement adds}

\JudgeR{} and \JudgeL{} answer related but different questions. \JudgeR{}
applies a predefined criterion that is appropriate for repeated regression
tests. \JudgeL{} asks whether the complete trace shows semantic compliance with
the attacker objective. \JudgeL{} labels 7.3\% of runs as partial, compared
with 2.0\% for \JudgeR{}. The additional cases include quoted canaries,
incomplete sink arguments, and traces in which the model's planning changed but
the final criterion was not met.

This disagreement is informative rather than incidental. The gap is largest
for attack methods that create ambiguous surface evidence. Obfuscation reaches
13.6\% attack success under \JudgeR{} in both carrier modes, but 9.1\% and
8.8\% under \JudgeL{}. In those cases, a sink event or target token can occur
without the trace establishing that the agent carried out the intended
objective. These are \emph{mechanical signals}: trace properties that a rule
can match directly, such as a sink call, a target token, a canary, or a
specific argument value. A mechanical signal demonstrates a behavioral change,
but it may not establish full semantic compliance. By contrast, fake completion
receives similar success rates from both evaluators. This pattern suggests that
A.I.G should retain the exact evidence used by \JudgeR{} and use \JudgeL{} to
prioritize traces for manual review, rather than collapsing the two outputs
into a single score.

\begin{keyfinding}
\textbf{Rule matches need semantic interpretation.} \code{obfuscation}
reaches 13.6\% \JudgeR{} full success in both carrier modes, but only 9.1\%
and 8.8\% under \JudgeL{}. Sink calls, target tokens, canaries, and argument
matches are useful evidence, but they do not always establish that the agent
completed the intended objective.
\end{keyfinding}

\section{Discussion}
\label{sec:discussion}

\subsection{Implications for DSH deployments}

The results do not reduce to a weakness in one prompt or one tool. The
relevant path includes external-content ingestion, carrier parsing, tool-result
serialization, model-visible session construction, planning, and tool-call
authorization. A deployment may have a capable sandbox or approval layer and
still be exposed if it passes untrusted content to the model without retaining
source information or constraining the action that follows.

\subsection{Controls suggested by the assessment}

\paragraph{Preserve provenance at the model boundary.}
Tool results should retain a source label, trust tier, and carrier type.
Normalization should expose hidden Unicode, metadata, and relevant
format-specific fields. System policy should state that content from an
untrusted source is data and cannot change the user's objective or authority.

\paragraph{Authorize sensitive sinks independently.}
Email, external HTTP submission, shell execution, file mutation, privilege
changes, and financial actions require controls that do not depend on the
model's interpretation of an external document. Those controls can include
allowlists, argument-level checks, data classification, and user approval.

\paragraph{Treat skills and integrations as code-adjacent assets.}
Skills, MCP integrations, tool descriptions, and workflow templates can all
shape model behavior. They should have ownership, provenance, version review,
and privilege restrictions. The skills result in this assessment makes that
point concrete.

\paragraph{Run the matrix after deployment changes.}
The relevant regression space includes wording, carrier, representation, tool
privilege, policy configuration, and model version. Teams should rerun selected
source-to-sink cases after changing prompts, tools, skills, parsers, model
providers, or authorization policy.

\section{Related work}
\label{sec:related}

\paragraph{Indirect prompt injection.}
Greshake et al.~\cite{greshake2023} demonstrated that instructions embedded
in content consumed by LLM-integrated applications can redirect a model away
from the user's request. The central distinction from direct prompt injection
is provenance: the malicious instruction arrives through an external artifact
that the application needs to process. Our assessment retains that distinction
by delivering the payload through a controlled DSH source tool rather than
placing it in the user message.

\paragraph{Agent-security benchmarks.}
InjecAgent~\cite{injecagent2024} evaluates indirect prompt injection in
tool-integrated agents, while AgentDojo~\cite{agentdojo2024} provides a dynamic
environment for evaluating attacks and defenses in agent workflows. These
benchmarks establish useful tasks and success conditions. Our goal is
different: we apply a broad carrier and wording matrix to a specific,
unmodified agent runtime, preserve its native session-event path, and retain
the resulting traces for both deterministic and semantic evaluation.

\paragraph{Prompt-boundary defenses.}
Spotlighting and data-marking approaches seek to distinguish trusted
instructions from untrusted content at the prompt boundary~\cite{hines2024}.
Boundary-awareness defenses in BIPIA~\cite{yi2025bipia} and structured-query
models in StruQ~\cite{chen2025struq} pursue the same separation through
different mechanisms. The present study does not compare these mitigations.
Instead, it identifies where a DSH deployment can apply them or related
source-aware controls: before a tool call, when processing a tool result, and
before a sensitive sink executes.

\paragraph{Agent assets, memory, and lifecycle controls.}
The elevated skills-channel result places reusable instructions and
integrations within the prompt-injection threat model. In practice, skills,
tool descriptions, retrieval connectors, and workflow templates should be
treated as reviewed assets with provenance and privilege constraints. This
observation is related to work on poisoned memory and knowledge bases for
agents~\cite{chen2024agentpoison,ying2026skilljack}, although our threat model focuses on
content that enters a live tool-result path rather than an optimized retrieval
backdoor. This distinction matters for deployment: both persistent assets and
transient external content need lifecycle controls.

\section{Conclusion}
\label{sec:conclusion}

This report used A.I.G to assess indirect prompt injection in DeepSeek Harness
under a controlled baseline configuration. The results show that the relevant
security boundary extends across the complete path from external content to a
sensitive action. Across 14,560 executions, fake-completion attacks achieved
17.0\% \JudgeL{} full success in text mode, while hidden Unicode and skills
attacks achieved 25.5\% and 16.0\% \JudgeR{} full success, respectively, in
file mode. These results reveal three distinct failure modes: workflow-like
language can alter task interpretation; file representations can change the
effective content received by the model; and reusable agent assets can carry
high-impact instructions across execution contexts.

The evaluation further demonstrates why aggregate attack-success rates are
insufficient on their own. Output-only objectives and sink-required objectives
can exhibit substantially different success rates, while rule-based evidence
does not necessarily imply complete semantic compliance. Source-to-sink
tracing makes these distinctions auditable by recording whether tainted content
reached the model, whether a sensitive tool was selected, and whether the
resulting action satisfied the task-specific success criterion. The analysis of
the DSH source code further identifies corresponding control points at
model-visible tool results, additional contexts, and pre-execution tool guards,
providing concrete locations for security hardening.

A.I.G is designed to make this style of assessment repeatable as agent systems
evolve. Its goal is not to assign a static security label to a framework, but
to provide developers and security teams with a practical means of testing the
composed systems they actually deploy, including their models, prompts, tools,
file parsers, skills, MCP integrations, authorization policies, and runtime
configurations. By combining controlled attack matrices, safe simulated sinks,
trace-level evidence, and complementary evaluators, A.I.G enables continuous
red teaming and regression testing of agent systems, extending security
assessment beyond one-time prompt checks to the full execution path through
which untrusted content can influence sensitive actions. We hope A.I.G can contribute to a more systematic approach to agent security, providing a practical foundation for understanding, measuring, and mitigating security risks across the agent execution lifecycle.

%% file: aig_appendix.tex
\clearpage
\appendix

\section{Reproducibility and Artifact Map}
\label{app:artifacts}

The report was derived from the local experiment artifacts accompanying this
technical report. The principal components are listed below.

\begin{table}[htbp]
\centering
\small
\caption{Key local artifacts used for the DSH case study.}
\begin{tabularx}{\textwidth}{lX}
\toprule
\textbf{Artifact} & \textbf{Role} \\
\midrule
\code{A.I.G assessment components} & Payload construction, taint routing, and trace normalization \\
\JudgeR{} / \JudgeL{} & Deterministic and semantic outcome evaluation \\
\code{full\_channel\_mode dataset} & 1,120-case text/file channel matrix \\
\code{experiment runner} & Matrix orchestration and online \JudgeR{} evaluation \\
\code{DSH runtime adapter} & Python-to-TypeScript harness bridge and trace mapping \\
\code{DSH driver} & Real DSH runtime initialization and session-event export \\
\code{controlled test plugin} & Source fixtures and tracked simulated sinks \\
\code{final trace corpus} & One normalized trace for each planned agent run \\
\code{offline judge corpus} & Trace-level \JudgeL{} verdicts \\
\bottomrule
\end{tabularx}
\end{table}

\section{Interpretation Notes}
\label{app:notes}

All sink tools in this evaluation are simulated. A recorded sink call
represents an attempted action by the agent; it does not send an email, execute
a command, transfer funds, or otherwise affect an external system. Percentages
are descriptive measurements for this controlled configuration, not claims of
universal vulnerability rates for DeepSeek Harness or any model provider.